\documentclass[]{aa}
\usepackage{graphicx}
\usepackage{color}
\usepackage{txfonts}

\newcommand{\beq}{\begin{equation}}
\newcommand{\eeq}{\end{equation}}
\newcommand{\beqa}{\begin{eqnarray}}
\newcommand{\eeqa}{\end{eqnarray}}
\newcommand{\mach}{M_{\mathrm{A}}}

\begin{document}

\title{Kelvin-Helmholtz instability of twisted magnetic flux tubes in the solar wind}

\author{T.~V.~Zaqarashvili\inst{1,3}, Z.~V\"or\"os\inst{1}, I.~Zhelyazkov\inst{2}}
 \institute{Space Research Institute, Austrian Academy of Sciences, 8042 Graz, Austria\\
             \email{teimuraz.zaqarashvili@oeaw.ac.at}
                \and
         Faculty of Physics, Sofia University, 5 James Bourchier Blvd., 1164 Sofia, Bulgaria\\
                \and
         Abastumani Astrophysical Observatory at Ilia State University, 3/5 Cholokashvili Avenue, 0162 Tbilisi, Georgia\\}

\date{Received / Accepted }

\abstract
{Solar wind plasma is supposed to be structured in magnetic flux tubes carried from the solar surface. Tangential velocity discontinuity near the boundaries of individual tubes may result in Kelvin-Helmholtz instability, which may contribute into the solar wind turbulence. While the axial magnetic field may stabilize the instability, a small twist in the magnetic field may allow to sub-Alfv\'enic motions to be unstable.
}
{We aim to study the Kelvin-Helmholtz instability of twisted magnetic flux tube in the solar wind with different configurations of external magnetic field.}
{We use magnetohydrodynamic equations in the cylindrical geometry and derive the dispersion equations governing the dynamics of twisted magnetic flux tube moving along its axis in the cases of untwisted and twisted external fields. Then we solve the dispersion equations analytically and numerically and found thresholds for Kelvin-Helmholtz instability in both cases of external field.}
{Both analytical and numerical solutions show that the Kelvin-Helmholtz instability is suppressed in the twisted tube by external axial magnetic field for sub-Alfv\'enic motions. However, even small twist in the external magnetic field allows the Kelvin-Helmholtz instability to be developed for any sub-Alfv\'enic motions. The unstable harmonics correspond to vortices with high azimuthal mode numbers, which are carried by the flow. }
{Twisted magnetic flux tubes can be unstable to Kelvin-Helmholtz instability when they move with small speed relative to main solar wind stream, then the Kelvin-Helmholtz vortices may significantly contribute into the solar wind turbulence.}


\titlerunning{Kelvin-Helmholtz instability of twisted tubes}

\authorrunning{Zaqarashvili et al.}

\maketitle
\section{Introduction}

The solar wind plasma is supposed to be composed of individual magnetic flux tubes which are carried from the solar atmosphere by the wind (Bruno et al.\ \cite{Bruno2001}, Borovsky \cite{Borovsky2008}).  Tangential velocity discontinuity at the tube surface due to the motion of tubes with regards to the solar wind stream may lead to the Kelvin-Helmholtz instability (KHI), which can be of importance as Kelvin-Helmholtz (KH) vortices may lead to the enhanced magnetohydrodynamic (MHD) turbulence.  Observations show that the velocity difference inside and outside the magnetic structures in the solar wind generally is not large, which means that the relative velocity of the tube and mean stream is sub-Alfv\'enic. KHI may develop for small velocity discontinuities in hydrodynamic flows (Drazin and Reid \cite{Drazin1981}), but a flow-aligned magnetic field stabilises sub-Alfv\'enic flows (Chandrasekhar \cite{Chandrasekhar1961}).  Therefore KHI will be suppressed in tubes moving with sub-Alfv\'enic speeds with regards to the solar wind.  On the other hand, a transverse magnetic field seems to have no effect on the instability (Sen \cite{Sen1963}, Ferrari et al.\ \cite{Ferrari1981}, Cohn \cite{cohn83}, Singh and Talwar \cite{Singh1994}), which means that the twisted magnetic tubes may become unstable to KHI even with sub-Alfv\'enic motions.

Solar wind flux tubes probably are ``fossil structures'' (i.e. they are carried from the solar atmosphere), then they may roughly keep the magnetic topology typical for tubes near the solar surface.  Complex photospheric motions may stretch and twist anchored magnetic field, which may lead to the consequent changes of topology at higher regions.  The observed rotation of sunspots (Khutsishvili et al.\ \cite{Khutsishvili1998}, Brown et al.\ \cite{Brown2003}, Yan and Qu \cite{Yan2007}, Zhang et al.\ \cite{Zhang2007}) may lead to the twisting of magnetic field above active regions, which can be observed as twisted loops in the corona (Srivastava et al.\ \cite{Srivastava2010}).  Recent observations of magnetic tornados (Wedemeyer-B\"ohm et al. \cite{Wedemeyer2012}, Su et al. \cite{Su2012}, Li et al.\ \cite{Li2012}) also strongly support the existence of twisted magnetic flux tubes on the Sun.  Newly emerged magnetic tubes can be also twisted during the rising phase through the convection zone (Moreno-Insertis and Emonet \cite{Moreno-Insertis1996}, Archontis et al.\ \cite{Archontis2004}, Murray and Hood \cite{Murray2008}, Hood et al.\ \cite{Hood2009}).  Therefore, solar magnetic tubes should have been twisted at photospheric, chromospheric and coronal levels. Helical magnetic flux rope can be also generated during the eruption of coronal mass ejections (CME's) and transported into the interplanetary space (Lynch et al. \cite{Lynch2004}). Original magnetic flux rope structure is supposed to be deformed during the transport through the heliosphere, but the flux rope still will keep its twisted nature (Manchester et al. \cite{Manchester2004}). Therefore, the solar wind magnetic tubes of all scales generally should be twisted. Twisted magnetic tubes are unstable to kink instability when the twist exceeds a critical value. The critical twist angle is $\sim\!\!70^\circ$, which means that the tubes twisted with a larger angle are unstable to the kink instability, therefore they probably can not reach 1 AU (Zaqarashvili et al. \cite{Zaqarashvili2013}). Twisted magnetic tubes can be observed by in situ vector magnetic field measurements in the solar wind considering force-free field model (Moldwin et al. \cite{Moldwin2000}, Feng et al. \cite{Feng2007}, Telloni et al. \cite{Telloni2012}), or by variation of total (magnetic + thermal) pressure (Zaqarashvili et al. \cite{Zaqarashvili2013}).

Here we study KHI of twisted magnetic flux tubes moving along their axes with regards to the mean solar wind stream.  The twist is assumed to be small enough, therefore the tubes are stable against kink instability. On the other hand, the harmonics with sufficiently high azimuthal mode number $m$ are always unstable to the KHI in the twisted magnetic tubes moving in nonmagnetic environment (Zaqarashvili et al.\ \cite{Zaqarashvili2010}).  However, the configuration of external magnetic field, which stabilizes KHI for sub-Alfv\'enic flows, is very important.  If the magnetic tubes move along the Parker spiral, then the external magnetic field is axial and KHI will be suppressed.  However, if the tubes move with angle to the Parker spiral, then the external magnetic field will have transverse component, which may allow KHI for sub-Alfv\'enic motions. In order to study the influence of transverse component of the external magnetic field on KHI, we consider a small twist in the external magnetic field in the cylindrical geometry, so that both, tube and external magnetic fields are stable for the kink instability.  In order to emphasis the role of transverse component of the external magnetic field, we consider external untwisted and twisted magnetic fields separately and derive KHI thresholds for both cases.

The paper is organized as follows. In Sect.~2 we consider the formulation of the problem and derive the solutions governing the plasma dynamics inside and outside the twisted tubes separately for untwisted and twisted external magnetic fields. In Sect.~3, we derive the dispersion equations of tube dynamics for untwisted and twisted external fields through boundary conditions at the tube surface.  In Sect.~4, we solve the dispersion equations both, analytically and numerically, and derive the instability thresholds for KHI.  Discussions of the problem and conclusion are presented in the last, fifth, section.

\section{Formulation of the problem and main solutions}

We consider a magnetic flux tube with radius $a$ embedded in a magnetized environment.  We use a cylindrical coordinate system $(r,\phi,z)$ and assume that the magnetic field has the following form: $\vec{B} = (0,B_{\phi}(r),B_{z}(r))$.  The unperturbed magnetic field and pressure satisfy the pressure balance condition
\begin{equation}
\label{balance0} {{\mathrm{d} }\over {\mathrm{d} r}}\left (p + {{B^2_{\phi} + B^2_{z}}\over {8\pi}}\right) = -{{B^2_{\phi}}\over {4 \pi r}}.
\end{equation}

We consider that the tube moves along the axial direction with regards to the surrounding medium, hence the flow profile inside the tube is $\vec{U} = (0,0,U)$.  In general, $U$ can be a function of $r$, but we consider the simplest homogeneous case.  No mass flow is considered outside the tube, which means that we are in the frame co-moving with the solar wind stream.

As the unperturbed parameters depend on the $r$ coordinate only, the perturbations can be Fourier analysed with $\exp [\mathrm{i}(m \phi + k_z z - \omega t)]$.  The equations governing the incompressible dynamics of the plasma are (Goossens
et al.\ \cite{Goossens1992})
\[
{{\mathrm{d}^2 p_\mathrm{t}}\over {\mathrm{d} r^2}} + \left[{{C_3}\over {r D}}{{\mathrm{d} }\over {\mathrm{d}
r}}\left ( {{r D}\over {C_3}}\right ) \right ]{{\mathrm{d} p_\mathrm{t}}\over {\mathrm{d} r}} +
\left[{{C_3}\over r D}{{\mathrm{d} }\over {\mathrm{d} r}}\left( {{r C_1}\over
{C_3}}\right) + {{C_2C_3-C^2_1}\over {D^2}}\right]p_\mathrm{t}
\]
\begin{equation}\label{general}
{}= 0,
\end{equation}
where
\[
D=\rho (\Omega^2 - \omega_\mathrm{A}^2),\qquad C_1=-{{2m B_{\phi}}\over {4 \pi
r^2}}\left ({m\over r}B_{\phi}+ k_z B_z\right ),
\]
\[
C_2=-\left ({m^2\over r^2}+ k_z^2\right ), \quad \; C_3=D^2 +
D{{2B_{\phi}}\over {4\pi}}{\mathrm{d}\over {\mathrm{d} r}}\left ({B_{\phi}\over
r}\right ) - {{4B^2_{\phi}}\over {4\pi r^2}}\rho \omega_\mathrm{A}^2,
\]
\begin{equation}\label{omegaA}
\omega_\mathrm{A} ={{\vec k \cdot \vec B}\over {\sqrt{4\pi \rho}}} =  {{1}\over {\sqrt{4\pi \rho}}}\left ({m\over r}B_{\phi}+ k_z
B_z\right )
\end{equation}
is the Alfv\'en frequency,
\begin{equation}\label{Omega}
\Omega=\omega-k_z U
\end{equation}
is the Doppler-shifted frequency and $p_\mathrm{t}$ is the total (hydrostatic $+$ magnetic) perturbed pressure.  Radial displacement $\xi_{r}$ is expressed through the total pressure as
\begin{equation}\label{eq:3}
\xi_{r}={D\over C_3}{{\mathrm{d} p_\mathrm{t}}\over {\mathrm{d} r}}+ {C_1\over C_3}p_{\mathrm{t}}.
\end{equation}

The solution to this equation depends on the magnetic field and density profiles.  Magnetic fields inside and outside the tube are denoted as $\vec B_\mathrm{i}$ and $\vec B_\mathrm{e}$ respectively, while the corresponding densities are $\rho_\mathrm{i}$ and $\rho_\mathrm{e}$.  To obtain the dispersion relation of oscillations we find the solutions inside and outside the tube and then merge the solutions at the tube boundary through boundary conditions.

\subsection{Solutions inside the tube}

We consider a magnetic flux tube with homogeneous density $\rho_i$ and uniform twist, i.e.,
\begin{equation}\label{magnetic-in}
\vec B_\mathrm{i} = (0,Ar,B_{\mathrm{i}z}),
\end{equation}
where $A$ is a constant.

In this case, Eq.~(\ref{general}) reduces to the modified Bessel equation
\begin{equation}\label{inside}
{{\mathrm{d}^2 p_\mathrm{t}}\over {\mathrm{d} r^2}}+{1\over r}{{\mathrm{d} p_\mathrm{t}}\over {\mathrm{d} r}}-\left
[{{m^2}\over {r^2}} + m^2_\mathrm{i}\right ]p_\mathrm{t} = 0,
\end{equation}
where
\begin{equation}\label{m0}
m^2_\mathrm{i} = k_z^2 \left[ 1-{{4A^2\omega^2_\mathrm{Ai}}\over {4\pi \rho_\mathrm{i} \left (\Omega^2 -
\omega^2_\mathrm{Ai}\right )^2}} \right],\qquad \omega_\mathrm{Ai} = {{mA + k_z
B_{\mathrm{i}z}}\over {\sqrt{4\pi \rho_\mathrm{i}}}}.
\end{equation}

A similar equation has been obtained by Dungey and Loughhead (\cite{Dungey1954}) and Bennett et al.\ (\cite{Bennett19}) in the
absence of flow, i.e., for $U = 0$.

\begin{figure}
   \includegraphics[width=8cm]{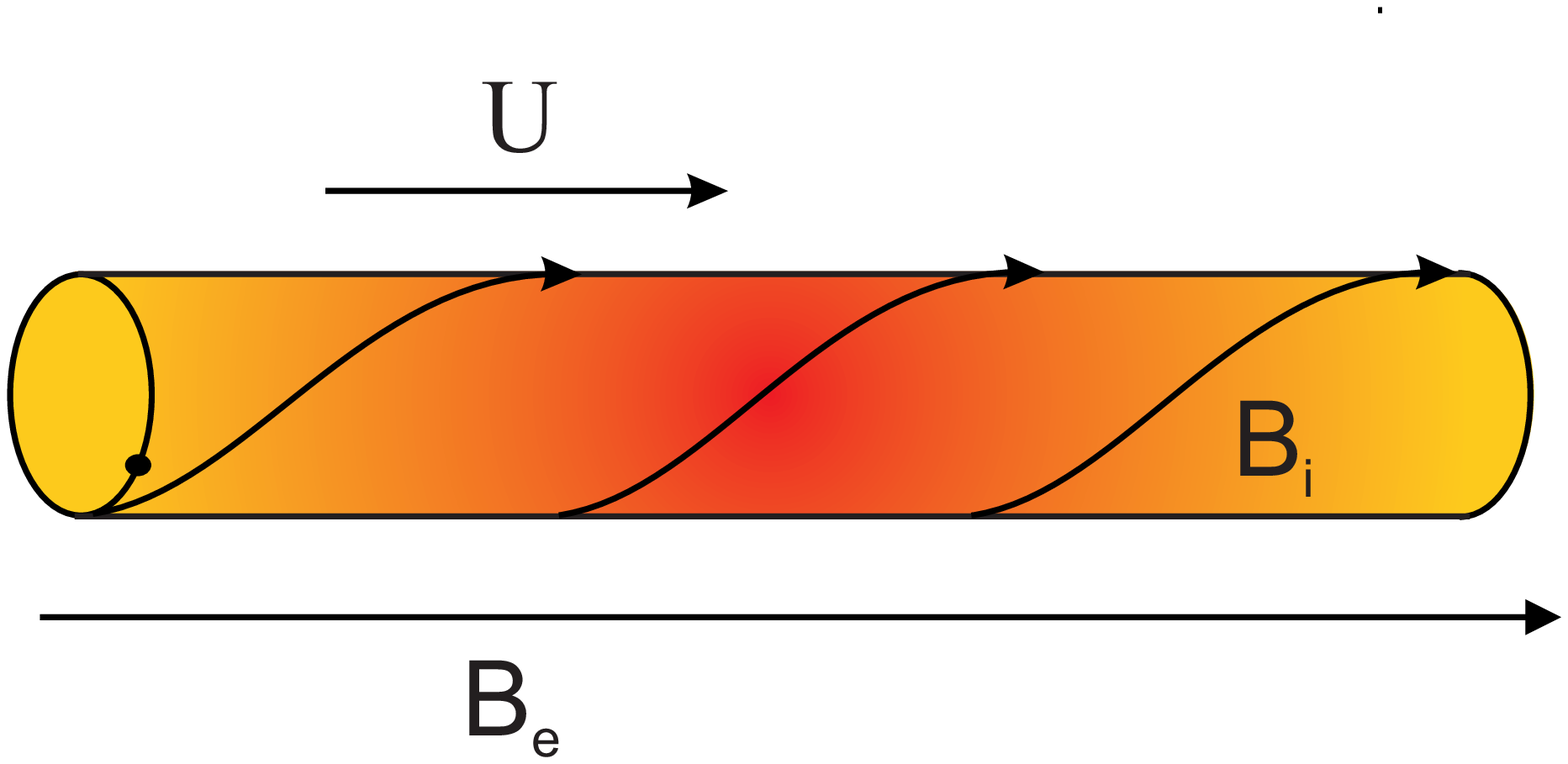}
   \includegraphics[width=8cm]{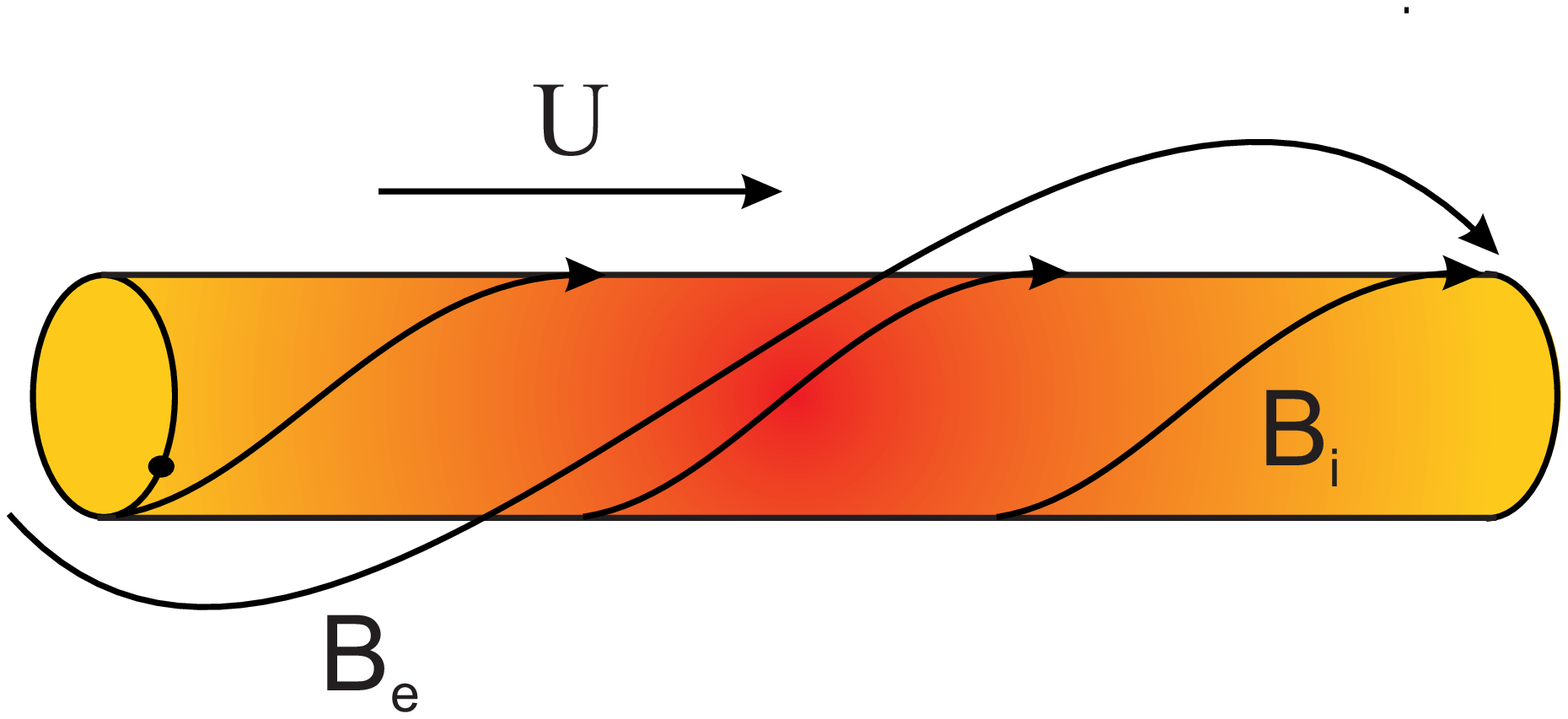}
      \caption{Twisted magnetic tube in two different configurations of external magnetic field: untwisted field (upper panel) and twisted field (lower panel). The tube moves along its axis with constant velocity, $U$, in both cases. }
         \label{dr_plot0}
   \end{figure}

The solution bounded at the tube axis is
\begin{equation}\label{ptin}
p_\mathrm{t} = a_\mathrm{i} I_m(m_\mathrm{i} r),
\end{equation}
where $I_m$ is the modified Bessel function of order $m$ and $a_\mathrm{i}$ is a constant.
Transverse displacement can be written using Eq.~(\ref{eq:3}) as
\begin{equation}\label{xiin}
\xi_{r}={a_\mathrm{i}\over r}\frac{\left (\Omega^2-\omega^2_\mathrm{Ai}\right )m_\mathrm{i}rI^{\prime}_m(m_\mathrm{i} r)-{2mA}\omega_\mathrm{Ai} I_m(m_\mathrm{i} r)/\!\!\sqrt{4\pi \rho_\mathrm{i}}} {\rho_\mathrm{i}\left (\Omega^2-\omega^2_\mathrm{Ai}\right )^2 - {4A^2}\omega^2_\mathrm{Ai}/4\pi},
\end{equation}
where the prime sign, $^{\prime}$, means a differentiation by the Bessel function argument.

\subsection{Solutions outside the tube}

Outside the tube we consider two different configurations of the magnetic field: untwisted (Fig.~\ref{dr_plot0}, upper panel) and twisted (Fig.~\ref{dr_plot0}, lower panel).

\subsubsection{Solution in the presence of external untwisted magnetic field}

In this case, we consider the homogeneous density, $\rho_\mathrm{e}$, and the homogenous external untwisted magnetic field of the form
\begin{equation}\label{magnetic-out}
{\vec B_\mathrm{e}} = (0,0,B_{\mathrm{e}z}).
\end{equation}
The total pressure perturbation outside the tube is governed by the same Bessel equation as Eq.~(\ref{inside}), but $m^2_\mathrm{i}$ is replaced by $k_z^2$. The solution bounded at infinity is
\begin{equation}\label{ptout1}
p_\mathrm{t} = a_\mathrm{e} K_m(k_z r),
\end{equation}
where $K_m$ is the modified Bessel function of order $m$ and $a_\mathrm{e}$ is a constant.

Transverse displacement can be written as
\begin{equation}\label{xiout1}
\xi_{r}={a_\mathrm{e}\over r}\frac{k_z rK^{\prime}_m(k_z r)} {\rho_{\mathrm{e}}\left (\omega^2-\omega^2_{\mathrm{Ae}}\right )},
\end{equation}
where, as before the prime sign, $^{\prime}$, means a differentiation by the Bessel function argument and
\begin{equation}\label{Ae}
\omega_\mathrm{Ae} = {{k_z
B_{\mathrm{e}z}}\over {\sqrt{4\pi \rho_\mathrm{e}}}}.
\end{equation}

\subsubsection{Solution in the presence of external twisted magnetic field}

In this case, we consider the external twisted magnetic field of the form
\begin{equation}\label{mag}
{\vec B_\mathrm{e}}=\left (0,B_{\mathrm{e} \phi}{a\over r}, B_{\mathrm{e}z}\left ({a\over r}\right )^2\right )
\end{equation}
and the density with the form $\rho = \rho_{\mathrm{e}}(a/r)^4$ so that the Alfv\'en frequency
\begin{equation}\label{Aenew}
\omega_\mathrm{Ae}={{mB_{\mathrm{e} \phi} + k_z aB_{\mathrm{e}z}}\over {\sqrt{4\pi\rho_{\mathrm{e}}a^2}}}
\end{equation}
is constant, which allows us to find an analytical solution of governing equation.  The total pressure perturbation outside the tube is governed by the Bessel-type equation
\begin{equation}\label{outside}
{{\mathrm{d}^2 p_\mathrm{t}}\over {\mathrm{d} r^2}}+{5\over r}{{\mathrm{d} p_\mathrm{t}}\over {\mathrm{d} r}}-\left ({{n^2}\over {r^2}}+ m^2_\mathrm{e} \right )p_\mathrm{t} = 0,
\end{equation}
where
\begin{equation}\label{n0}
n^2 = m^2 - {{4 m^2 B^2_{\mathrm{e} \phi}}\over {4\pi\rho_{\mathrm{e}} a^2 \left (\omega^2-\omega^2_\mathrm{Ae}\right )}}+{{8 m B_{\mathrm{e} \phi}\omega_\mathrm{Ae}}\over {\sqrt{4\pi\rho_{\mathrm{e}}} a \left (\omega^2-\omega^2_\mathrm{Ae}\right )}},
\end{equation}
and
\begin{equation}\label{me}
m^2_\mathrm{e} = k_z^2\left (1-{{4B^2_{\mathrm{e} \phi}\omega^2}\over {4\pi\rho_{\mathrm{e}}\left (\omega^2 - \omega^2_\mathrm{Ae}\right )^2 a^2}}\right ).
\end{equation}

A solution to this equation bounded at infinity is
\begin{equation}\label{ptout2}
p_\mathrm{t} = a_\mathrm{e} {{a^2}\over {r^2}} K_{\nu}(m_\mathrm{e}r),
\end{equation}
where
\begin{equation}\label{nu}
\nu = \sqrt{4 + m^2-{{4 m^2 B^2_{\mathrm{e} \phi}}\over {4\pi\rho_{\mathrm{e}} a^2 \left (\omega^2-\omega^2_\mathrm{Ae}\right )}}+{{8 m B_{\mathrm{e} \phi}\omega_\mathrm{Ae}}\over {\sqrt{4\pi\rho_{\mathrm{e}}} a \left (\omega^2-\omega^2_\mathrm{Ae}\right )}}},
\end{equation}
and $a_\mathrm{e}$ is a constant.

Transverse displacement can be written as
\[
\xi_{r} = a_\mathrm{e}\frac{r\left (\omega^2 - \omega^2_\mathrm{Ae}\right )(m_\mathrm{e}r)K^{\prime}_{\nu}(m_\mathrm{e}r)}{a^2 \rho_{\mathrm{e}}\left (\omega^2-\omega^2_\mathrm{Ae}\right )^2 - {{4B^2_{\mathrm{e} \phi}}}\omega^2/4\pi}
\]
\begin{equation}\label{xiout2}
{}-a_\mathrm{e}{r\over a}\left (\frac{2a\left (\omega^2 - \omega^2_\mathrm{Ae}\right )+{{2mB_{\mathrm{e} \phi}}}\omega_\mathrm{Ae}/\sqrt{4\pi\rho_{\mathrm{e}}}}{a^2\rho_{\mathrm{e}}\left (\omega^2-\omega^2_\mathrm{Ae}\right )^2-{{4B^2_{\mathrm{e} \phi}}}\omega^2/4\pi} \right )K_{\nu}(m_\mathrm{e}r).
\end{equation}

\section{Dispersion equations}

Merging the solutions, i.e., Eqs.~(\ref{ptin}), (\ref{xiin}), (\ref{ptout1}), (\ref{xiout1}), (\ref{ptout2}), and (\ref{xiout2}), at the tube surface, $r = a$, leads to the dispersion equations governing the dynamics of magnetic tube.  In the following we always consider positive $k_z$.  The boundary conditions at the tube surface are the continuity of Lagrangian displacement and total Lagrangian pressure (Dungey and Loughhead \cite{Dungey1954}, Bennett et al.\ \cite{Bennett19}), i.e.,
\begin{equation}\label{boundary1}
[\xi_r]_{a} = 0
\end{equation}
and
\begin{equation}\label{boundary2}
\left [p_\mathrm{t} - {{B^2_{\phi}}\over {4\pi a}}\xi_r \right]_{a} = 0.
\end{equation}

Using these conditions we can derive the dispersion equations governing the oscillations of moving twisted magnetic tube in both, untwisted and twisted external magnetic fields.

\subsection{Dispersion equation for the external untwisted magnetic field}

Using Eqs.~(\ref{ptin})--(\ref{xiin}) and Eqs.~(\ref{ptout1})--(\ref{xiout1}) the following dispersion relation is obtained
\[
{{([\omega-k_z U]^2-\omega^2_\mathrm{Ai})F_m(m_\mathrm{i}a)-2mA\omega_\mathrm{Ai}/\!\sqrt{4 \pi \rho_\mathrm{i}}}\over {\rho_\mathrm{i}([\omega-k_z U]^2-\omega^2_\mathrm{Ai})^2-4A^2\omega^2_\mathrm{Ai}/4 \pi}}
\]
\begin{equation}
\label{dispersion1}
{}={{P_{m}(k_z a)}\over {\rho_{\mathrm{e}}(\omega^2-\omega^2_\mathrm{Ae})+A^2P_{m}(k_z a)/4 \pi }},
\end{equation}
where
\[
F_{m}(m_\mathrm{i}a)={{m_\mathrm{i}aI^{\prime}_{m}(m_\mathrm{i}a)}\over {I_{m}(m_\mathrm{i}a)}},\qquad P_{m}(k_z a)={{k_z aK^{\prime}_{m}(k_z a)}\over {K_{m}(k_z a)}}.
\]
This equation is the same as Eq.~(13) in Zhelyazkov and Zaqarashvili (\cite{Zhelyazkov2012}) with different notations.

\subsection{Dispersion equation for the external twisted magnetic field}

Using Eqs.~(\ref{ptin})--(\ref{xiin}) and Eqs.~(\ref{ptout2})--(\ref{xiout2}) the following dispersion relation is obtained
\[
{{\left ([\omega-k_z U]^2-\omega^2_\mathrm{Ai} \right )F_m(m_\mathrm{i}a)-2mA\omega_\mathrm{Ai}/\!\sqrt{4 \pi \rho_\mathrm{i}}}\over {\rho_\mathrm{i}\left ([\omega-k_z U]^2 - \omega^2_\mathrm{Ai}\right )^2 - 4A^2\omega^2_\mathrm{Ai}/4 \pi}}
\]
\begin{equation}
\label{dispersion2}
{}={{a^2\left (\omega^2-\omega^2_\mathrm{Ae}\right )Q_{\nu}(m_\mathrm{e}a)-G}\over {L - H\left [a^2\left (\omega^2-\omega^2_\mathrm{Ae}\right )Q_{\nu}(m_\mathrm{e}a) - G\right ]}},
\end{equation}
where
\[
Q_{\nu}(m_\mathrm{e}a) = {{m_\mathrm{e}aK^{\prime}_{\nu}(m_\mathrm{e}a)}\over {K_{\nu}(m_\mathrm{e}a)}},\quad L = a^2 \rho_{\mathrm{e}}\left (\omega^2-\omega^2_\mathrm{Ae}\right )^2-{{4B^2_{\mathrm{e} \phi}\omega^2}\over {4\pi}},
\]
\[
H = {{B^2_{\mathrm{e} \phi}}\over {4\pi a^2}} -{{A^2}\over {4\pi}}, \quad G = 2a^2 \left (\omega^2-\omega^2_\mathrm{Ae}\right )+{{2maB_{\mathrm{e} \phi}\omega_\mathrm{Ae}}\over {\sqrt{4\pi\rho_{\mathrm{e}}}}} .
\]

\section{Instability criteria}

Dispersion equations (\ref{dispersion1}) and (\ref{dispersion2}) govern the dynamics of twisted tubes moving in external untwisted and twisted magnetic fields, respectively. If the frequency, $\omega$, is complex value, then it indicates an instability process in the system; real part corresponds to the oscillation and imaginary part corresponds to the growth rate of instability.  Two types of instability may develop in moving twisted tubes: kink instability due to the twist and KHI due to the tangential discontinuity of flow at the tube surface.  However, only KHI remains for weakly twisted tubes.  Therefore, the condition of complex frequency in Eqs.~(\ref{dispersion1}) and (\ref{dispersion2}) determines the criterion of KHI in weakly twisted tubes.

Equations (\ref{dispersion1}) and (\ref{dispersion2}) are transcendental equations with Bessel functions of complex argument and complex order (for Eq.~\ref{dispersion2}). We first solve the dispersion equations analytically using long wavelength approximation and obtain corresponding analytical instability criteria, then we solve the dispersion equations numerically.

\subsection{Instability criterion of twisted magnetic tubes embedded in untwisted external magnetic field}

The long wave length approximation, $k_z a \ll 1$, yields $m_\mathrm{i}a \ll 1$, therefore we have
\[
F_m(m_\mathrm{i}a)=\frac{m_\mathrm{i} aI^{\prime}_m(m_\mathrm{i}a)}{I_m(m_\mathrm{i}a)}\approx |m|
\]
and
\[
P_m(ka) = \frac{k_z aK^{\prime}_m(k_z a)}{K_m(k_z a)}\approx -|m|.
\]
Then Eq.~(\ref{dispersion1}) gives the polynomial dispersion relation
\[
\omega^2 -{{2\rho_\mathrm{i} k_z U}\over {\rho_\mathrm{i} + \rho_{\mathrm{e}}}}\omega + {{\rho_\mathrm{i}}\over {\rho_\mathrm{i} + \rho_{\mathrm{e}}}}k_z^2U^2 - {{\rho_\mathrm{i}}\over {\rho_\mathrm{i} + \rho_{\mathrm{e}}}}\omega^2_\mathrm{Ai}-
{{\rho_{\mathrm{e}}}\over {\rho_\mathrm{i} + \rho_{\mathrm{e}}}}\omega^2_\mathrm{Ae}
\]
\begin{equation}\label{disp1}
{}-{{A^2|m|}\over {4\pi(\rho_\mathrm{i} + \rho_{\mathrm{e}})}} + {{2A\omega_\mathrm{Ai}\sqrt{4\pi\rho_\mathrm{i}}}\over {4\pi(\rho_\mathrm{i} + \rho_{\mathrm{e}})}} = 0.
\end{equation}

We consider perturbations with wave vector nearly perpendicular to the magnetic field, i.e., ${\vec k}\cdot{\vec B}\approx 0$, which seem to be most unstable ones (Pataraya and Zaqarashvili
\cite{Pataraya1995}).  These modes are pure vortices in the incompressible limit, therefore they have strongest growth rate due to KHI.  In cylindrical coordinates, this condition is expressed inside the tube as
\begin{equation}\label{kB}
m \approx - {{k_z B_{\mathrm{i}z}}\over {A}}.
\end{equation}
Then, Eq.~(\ref{disp1}) is simplified and we have
\begin{equation}\label{disp11}
\omega^2 -{{2\rho_\mathrm{i} k_z U}\over {\rho_\mathrm{i} + \rho_{\mathrm{e}}}}\omega + {{\rho_\mathrm{i} k_z^2U^2}\over {\rho_\mathrm{i} + \rho_{\mathrm{e}}}} -
{{\rho_{\mathrm{e}}\omega^2_\mathrm{Ae}}\over {\rho_\mathrm{i} + \rho_\mathrm{e}}}-{{A^2|m|}\over {4\pi(\rho_\mathrm{i} + \rho_\mathrm{e})}} = 0.
\end{equation}
\begin{figure}
   \includegraphics[width=8.85cm]{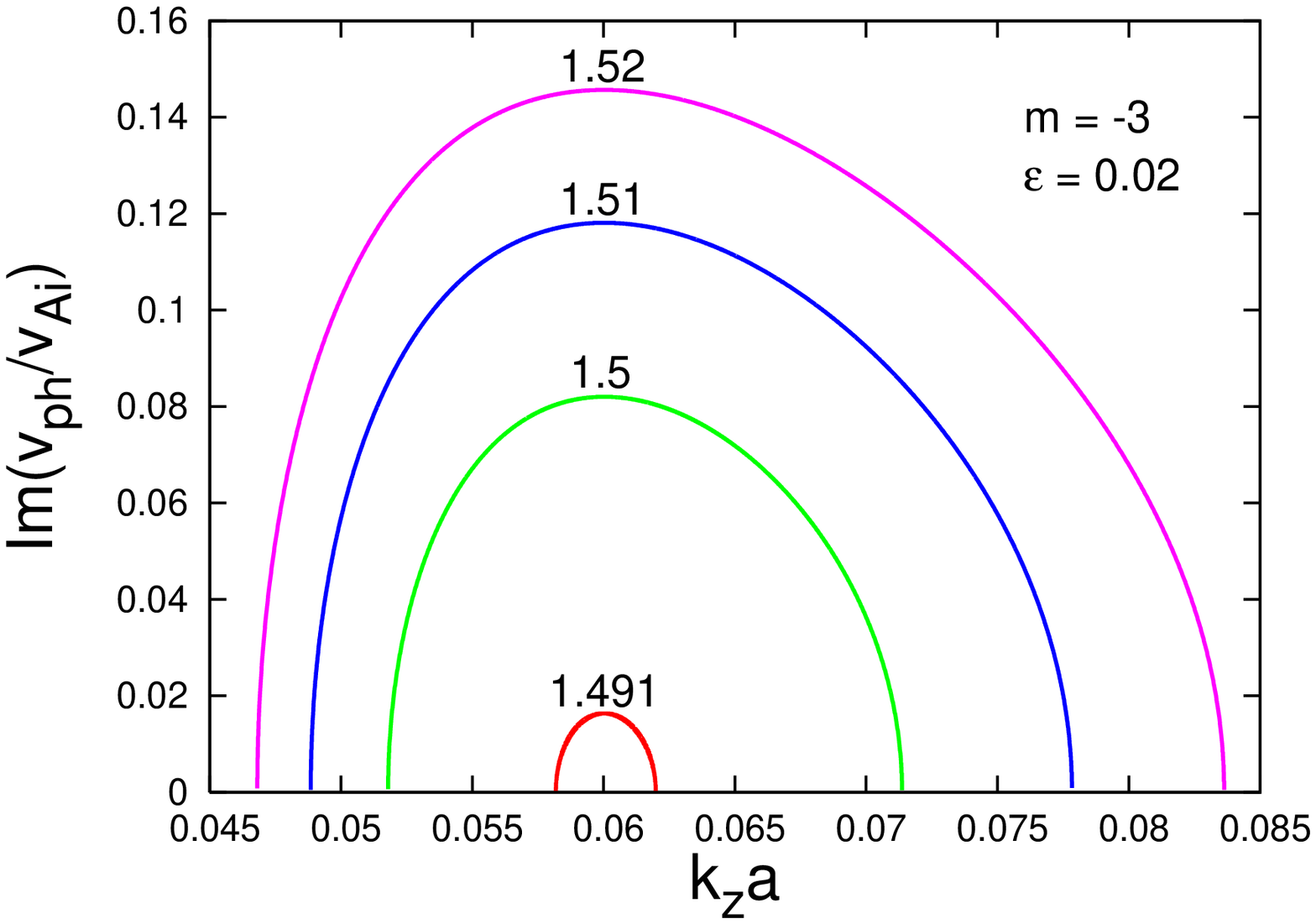}
   \\
   \\
   \includegraphics[width=8.85cm]{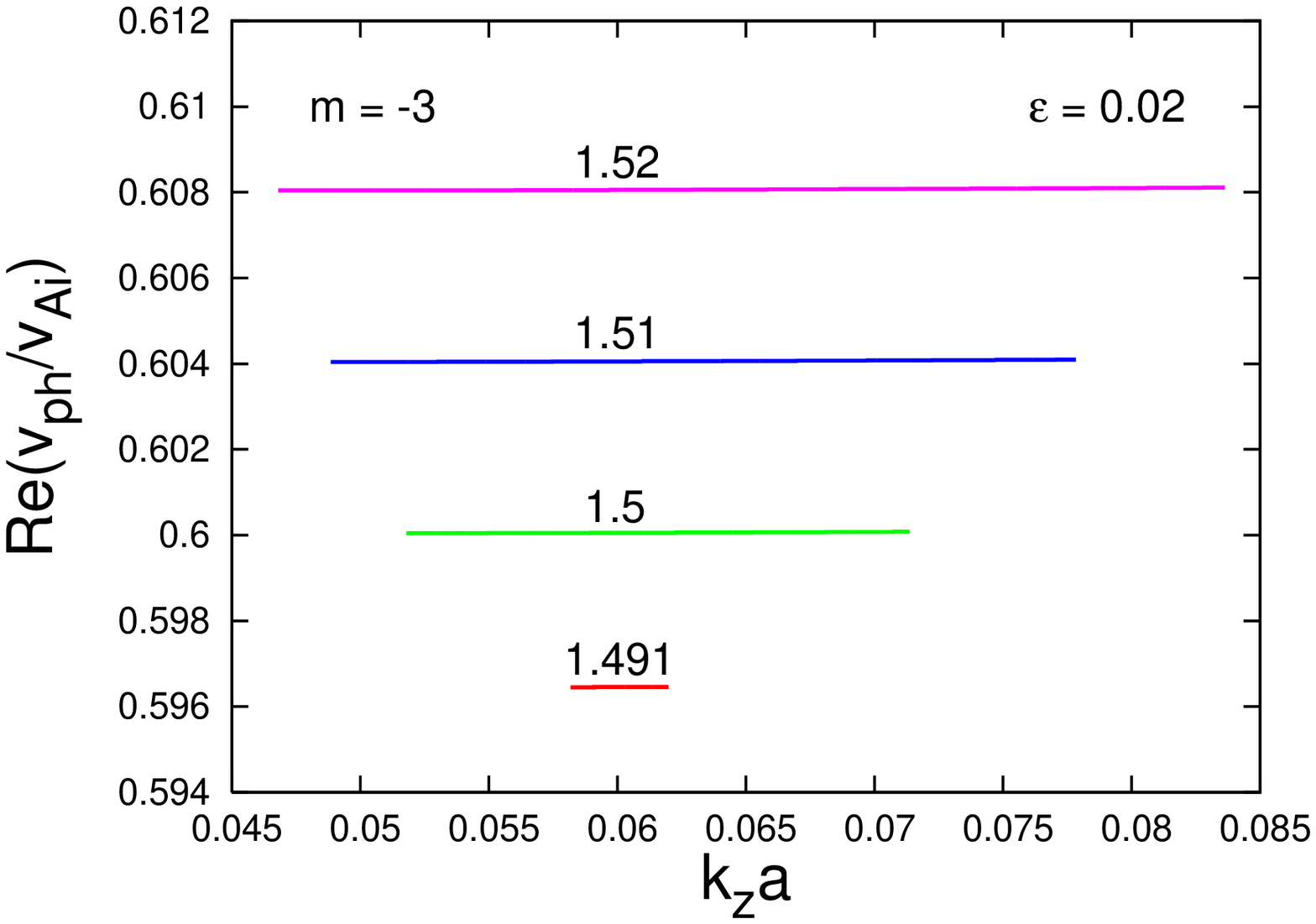}
      \caption{The real (lower panel) and imaginary (upper panel) parts of normalized phase speed, $v_{\mathrm{ph}}/v_{\mathrm{Ai}} = \omega/(k_z v_{\mathrm{Ai}})$, vs normalized wave number $k_z a$ of $m = -3$ unstable harmonics for external untwisted magnetic field (after numerical solution of dispersion equation~(\ref{dispersion1})). Red, green, blue, and magenta lines correspond to Alfv\'en Mach numbers $M_\mathrm{A} = 1.491, 1.5, 1.51$, and $1.52$, respectively. Here we assume the following parameters: $\rho_\mathrm{i}/\rho_\mathrm{e} = 0.67$, $\varepsilon = B_{\mathrm{i} \phi}/B_{\mathrm{i}z}= Aa/B_{\mathrm{i}z} = 0.02$, and $B_{\mathrm{i}z}/B_{\mathrm{e}z} = 1$.}
         \label{dr_plot1}
   \end{figure}

Kelvin-Helmholtz instability yields the complex frequency, $\omega$, therefore Eq. (\ref{disp11}) gives the instability criterion
as
\begin{equation}\label{KH1}
|m|\mach^2> \left (1 + {\rho_\mathrm{i}\over \rho_\mathrm{e}}\right )\left (|m|{{B^2_{\mathrm{e} z}}\over {B^2_{\mathrm{i}z}}}+1\right ),
\end{equation}
where
\begin{equation}\label{Mach}
\mach={U\over v_{\mathrm{Ai}}}
\end{equation}
is the Alfv\'en Mach number and $v_\mathrm{Ai}=B_{\mathrm{i}z}/\!\sqrt{4\pi\rho_\mathrm{i}}$ is the Alfv\'en speed inside the tube.

This criterion means that only super-Alfv\'enic flows are unstable to KHI in the case of external axial magnetic field. In principle, sub-Alfv\'enic flows can be also unstable in the case of weak external magnetic field.  For $B_{\mathrm{e}z} = 0$ Eq.~(\ref{KH1}) is transformed into the criterion of KHI in the twisted magnetic tube with nonmagnetic environment (Eq.~(28) in Zaqarashvili et al.\ \cite{Zaqarashvili2010}).  But, if the internal and external magnetic fields have similar strengths then KHI only starts for super-Alfv\'enic motions.

\subsection{Instability criterion of twisted magnetic tubes embedded in twisted external magnetic field}

The long wave length approximation, $k_z a \ll 1$, yields $m_\mathrm{i}a \ll 1$ and $m_\mathrm{e}a \ll 1$, therefore we have
\[
F_m(m_\mathrm{i}a)=\frac{m_\mathrm{i}aI^{\prime}_m(m_\mathrm{i}a)}{I_m(m_\mathrm{i}a)}\approx |m|
\]
and
\[
Q_{\nu}(m_\mathrm{e}a)={{m_\mathrm{e}aK^{'}_{\nu}(m_\mathrm{e}a)}\over {K_{\nu}(m_\mathrm{e}a)}}\approx -|\nu|.
\]
We assume that the ratio of azimuthal components of external and internal magnetic field is small, i.e., $B_{\mathrm{e} \phi}/(aA)\ll 1$ (this yields $|\nu| \approx |m|$) and consider the perturbations with ${\vec k}\cdot{\vec B_\mathrm{e}}\approx 0$.  Then Eq.~(\ref{dispersion2}) gives the polynomial dispersion relation
\[
\left (1+{{|m|}\over {2+|m|}}{\rho_\mathrm{e}\over \rho_\mathrm{i}}\right )\omega^2 -2 k_z U\omega + k_z^2U^2 - \omega^2_\mathrm{Ai}+{{2A\omega_\mathrm{Ai}}\over {\sqrt{4\pi\rho_\mathrm{i}}}}-
\]
\begin{equation}\label{disp2}
-{{A^2m}\over {4\pi\rho_\mathrm{i}}} = 0.
\end{equation}

We can further simplify Eq.~(\ref{disp2}) considering ${\vec k}\cdot{\vec B_\mathrm{i}}\approx 0$, which gives
\begin{equation}\label{disp3}
\left (1+{|m|\over {2+|m|}}{\rho_\mathrm{e}\over \rho_\mathrm{i}}\right )\omega^2 -2 k_z U\omega + k_z^2 U^2-{{A^2m}\over {4\pi\rho_\mathrm{i}}} = 0.
\end{equation}
Then the instability criterion is
\begin{equation}\label{KH2}
|m|\mach^2> 1 + {{2+|m|}\over {|m|}}{\rho_\mathrm{i}\over \rho_\mathrm{e}}.
\end{equation}

Eq.~(\ref{KH2}) shows that the harmonics with sufficiently high $m$ are unstable for any value of $\mach$.  The threshold of KHI decreases for higher $m$. For example, the threshold Mach number for $m = -1$ harmonics is $M_\mathrm{A} \approx 1.73$, while for $m = -4$ harmonics it is reduced to $M_\mathrm{A} \approx 0.707$ (${\rho_\mathrm{i}/\rho_\mathrm{e}} = 0.67$ is assumed during the estimation).  Therefore, the motion of the tube with the speed of $0.71 v_\mathrm{Ai}$ leads to the instability of harmonics with azimuthal mode numbers $|m| \geqslant 4$.  The tubes with lower speed will be unstable to higher $m$ harmonics.  The criterion of Kelvin-Helmholtz instability Eq.~(\ref{KH2}) is similar to the case of twisted magnetic tube in nonmagnetic environment (Zaqarashvili et al.\ \cite{Zaqarashvili2010}).

\begin{figure}
   \includegraphics[width=9cm]{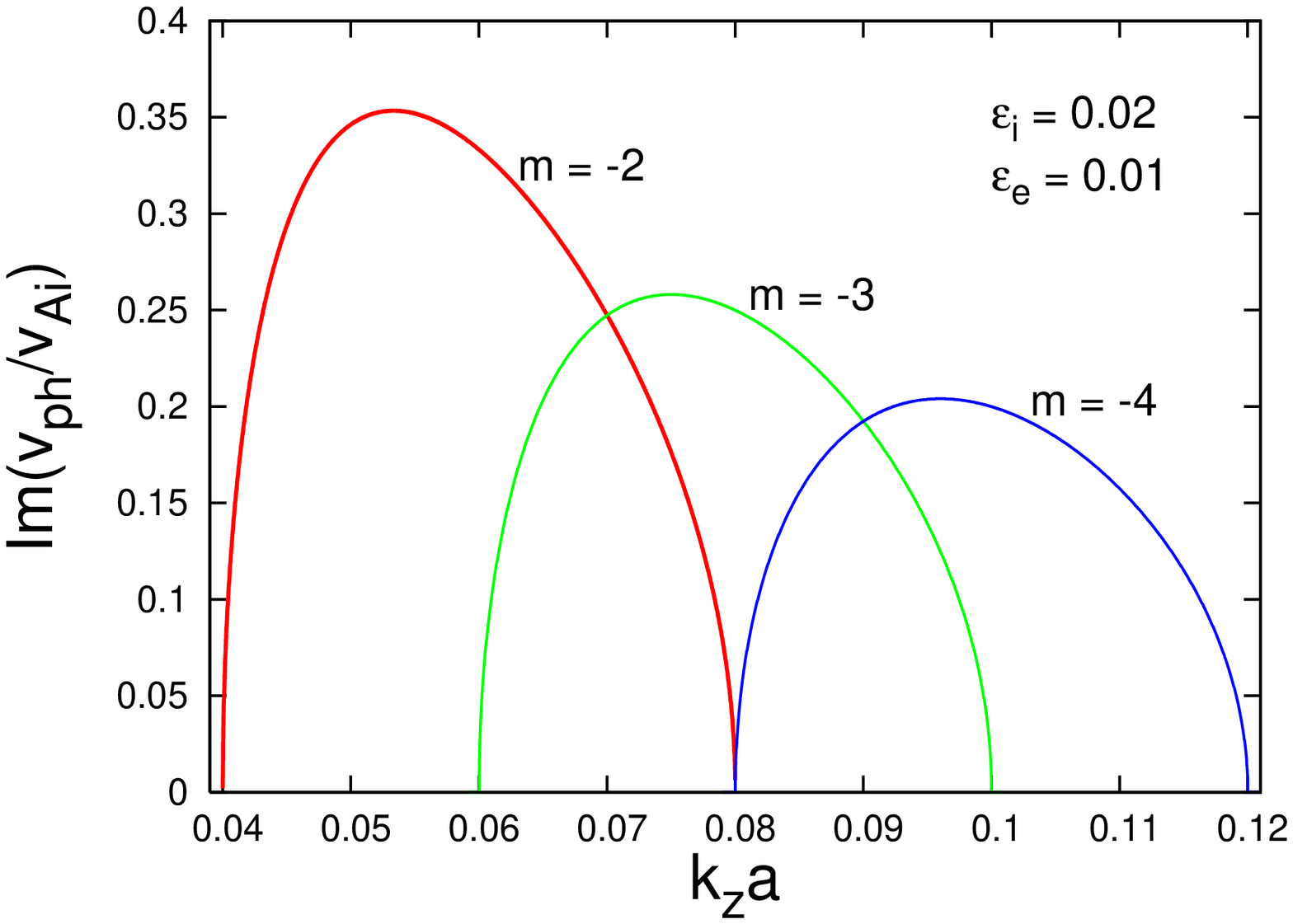}
   \\
   \\
   \includegraphics[width=9cm]{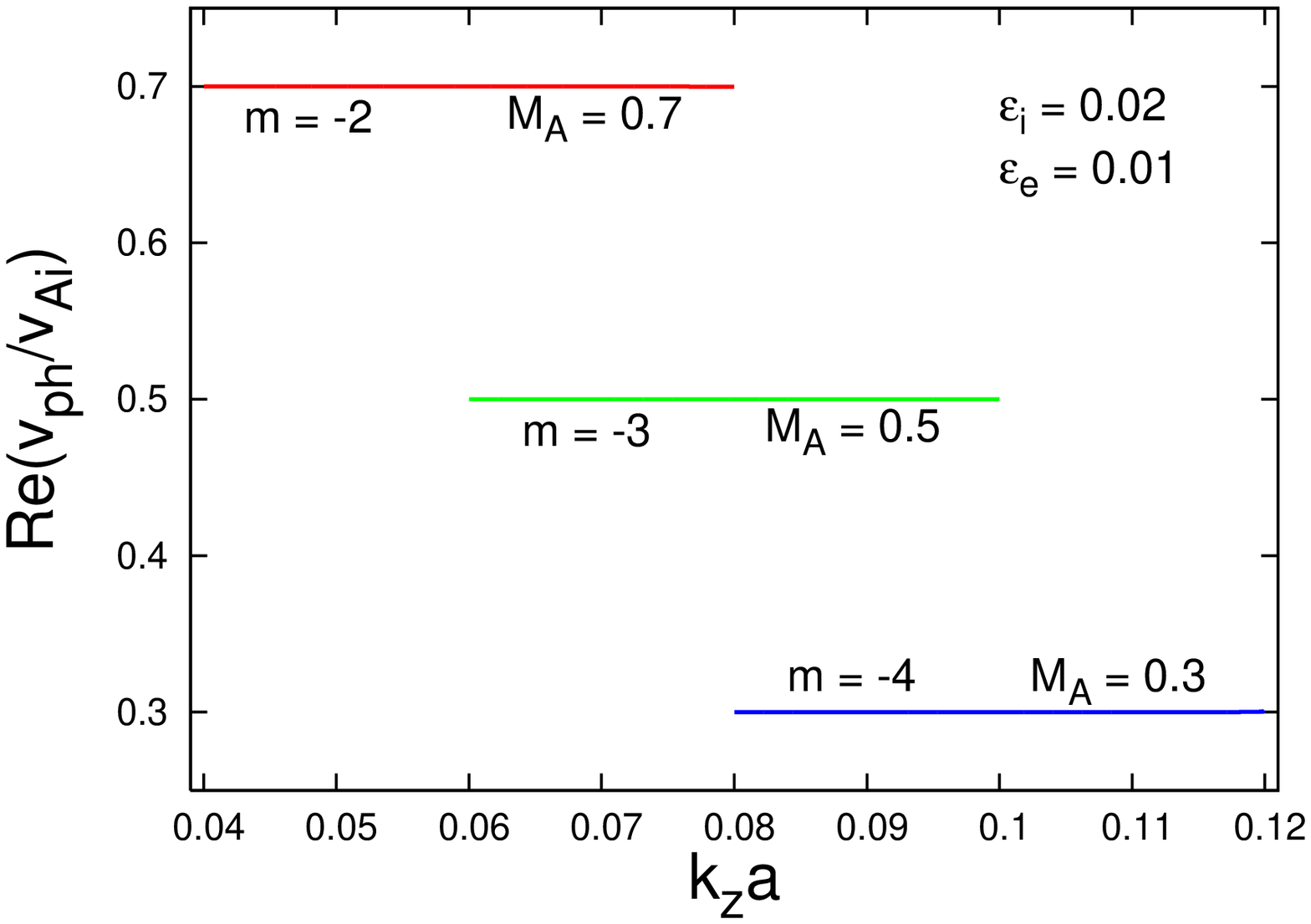}
      \caption{The real (lower panel) and imaginary (upper panel) parts of normalized phase speed, $v_\mathrm{ph}/v_\mathrm{Ai}=\omega/(k_z v_\mathrm{Ai})$, vs normalized wave number $k_z a$ of unstable harmonics for different values of Alfv\'en Mach number (after numerical solution of dispersion equation Eq.~(\ref{dispersion2})). Red, green, and blue lines indicate the phase speeds and the growth rates of $m = -2$ harmonics for Alfv\'en Mach number $M_\mathrm{A} = 0.7$, $m = -3$ harmonics for Alfv\'en Mach number $M_\mathrm{A} = 0.5$ and $m = -4$ harmonics for Alfv\'en Mach number $M_\mathrm{A} = 0.3$, respectively.  Here we assume the following parameters: $\rho_\mathrm{i}/\rho_\mathrm{e}=0.67$, $\varepsilon_\mathrm{i} = B_{\mathrm{i} \phi}/B_{\mathrm{i}z} = Aa/B_{\mathrm{i}z} = 0.02$, $\varepsilon_\mathrm{e} = B_{\mathrm{e} \phi}/B_{\mathrm{e}z} = 0.01$, $B_{\mathrm{i}z}/B_{\mathrm{e}z} = 1$.}
         \label{dr_plot}
   \end{figure}

\section{Numerical solutions of dispersion equations}

In order to check the analytical solutions, we solved the dispersion equations (\ref{dispersion1}) and (\ref{dispersion2}) numerically.  Numerical solution of dispersion Eq.~(\ref{dispersion1}) shows that all harmonics are stable for sub-Alfv\'enic flows, $M_\mathrm{A} < 1$.  Therefore, only super-Alfv\'enic flows, $M_\mathrm{A} > 1$, are unstable in the case of external untwisted magnetic field.  Fig.~\ref{dr_plot1} shows $m = -3$ unstable harmonics for different values of $M_\mathrm{A}$ after solution of Eq.~(\ref{dispersion1}).  It is seen that the critical Alfv\'en Mach number equals to $1.491$ for $m = -3$ harmonics.  Analytically estimated critical Mach number (from Eq.~(\ref{KH1})) is $M_\mathrm{A} \approx 1.4907$ for $m = -3$ harmonics.  Hence there is a very good agreement between analytical and numerical values. Normalized wave number of unstable harmonics is around $k_z a \approx 0.06$ in Fig.~\ref{dr_plot1}, which confirms that the unstable harmonics correspond to the condition ${\vec k}\cdot{\vec B_\mathrm{i}}\approx 0$, which implies $k_z a \approx -m \varepsilon_\mathrm{i}$, where $\varepsilon_\mathrm{i} = B_{\mathrm{i} \phi}/B_{\mathrm{i}z}=Aa/B_{\mathrm{i}z}$.  Both, numerical and analytical solutions to Eq.~(\ref{dispersion1}) agree with the well-known result that the flow-aligned magnetic field stabilizes KHI.

On the other hand, the numerical solution to Eqs.~(\ref{dispersion2}) confirms that there are unstable harmonics with sufficiently high azimuthal mode number $m$ for any value of Alfv\'en Mach number in the case of twisted tube with twisted external magnetic field.  Fig.~\ref{dr_plot} shows the unstable harmonics for different values of $M_\mathrm{A}$ (after solution of dispersion equation Eq.~(\ref{dispersion2})).  Alfv\'en Mach number $M_\mathrm{A} = 0.7$ yields the lowest azimuthal wave number of the unstable harmonics as $m = -2$ and the longitudinal wave number $k_z a$ is located in the interval of $0.04$--$0.08$. Therefore, harmonics with $|m|\geqslant 2$ are unstable for $M_\mathrm{A} = 0.7$. Alfv\'en Mach number $M_\mathrm{A} = 0.5$ yields the lowest azimuthal wave number as $m = -3$ and the longitudinal wave numbers lay inside the interval of $0.06$--$0.1$.  In the same way, Alfv\'en Mach number $M_\mathrm{A} = 0.3$ yields the lowest azimuthal wave number and the longitudinal wave number interval as $m = -4$ and $0.08$--$0.12$, respectively.  Hence, higher $m$ harmonics yield lower Alfv\'en Mach number in order to become unstable.  Numerically estimated thresholds yield lower value as compared to the analytical instability criterion Eq.~(\ref{KH2}).  For example, the harmonics with $m = -3$ yield the instability threshold of $M_A \approx 0.8$ from Eq.~(\ref{KH2}), while the numerically obtained threshold is $M_\mathrm{A} \approx 0.5$.  Numerical solutions again confirm that the unstable harmonics correspond to the condition ${\vec k}\cdot{\vec B_\mathrm{i}}\approx 0$.  We found that the unstable harmonics with $m = -3$ start to be unstable for $\varepsilon_\mathrm{i} = 0.02$ when $k_z a=0.06$ as it is expected (see Fig.~\ref{dr_plot}).

Note that the harmonics with negative $m$ and positive $\varepsilon_\mathrm{i}$ have identical properties to the harmonics with positive $m$ and negative $\varepsilon_\mathrm{i}$.  Fig.~\ref{dr_plot1} and Fig.~\ref{dr_plot} show that the phase speed of unstable harmonics corresponds to the flow speed $U$, which is the speed of magnetic tube with regards to the solar wind.  It is an expected result as ${\vec k}\cdot{\vec B_\mathrm{i}}\approx 0$ condition corresponds to pure vortex solutions.  Then the vortices are carried by the flow and consequently the phase speed of perturbations equals the flow speed.

\section{Discussion}

Recent observations of KH vortices in solar prominences (Berger et al.\ \cite{Berger2010}, Ryutova et al.\ \cite{Ryutova2010}) and at boundaries of rising CMEs (Foullon et al.\ \cite{Foullon2011}, Ofman and Thompson \cite{Ofman2011}, M\"ostl et al.\ \cite{Mostl2013}) increased the interest towards KHI in magnetic flux tubes.  KHI has been studied in the presence of kink oscillations in coronal loops (Terradas et al.\ \cite{Terradas2008}, Soler et al.\ \cite{Soler2010}), in twisted magnetic flux tubes with nonmagnetic environment (Zaqarashvili et al.\ 2010), magnetic tubes with partially ionized plasmas (Soler et al.\ \cite{Soler2012}), in spicules (Zhelyazkov \cite{Zhelyazkov2012a}) and soft X-ray jets (Zhelyazkov \cite{Zhelyazkov2012b}), as well as in photospheric tubes (Zhelyazkov and Zaqarashvili \cite{Zhelyazkov2012}).

KH vortices can be considered as one of important sources for MHD turbulence in the solar wind.  KHI can be developed by velocity discontinuity at boundaries of magnetic flux tubes owing to the relative motion of the tubes with regards to the solar wind or neighboring tubes.  However, flow-aligned magnetic field may suppress KHI for typical velocity jump at boundaries of observed magnetic structures in the solar wind, which is generally sub-Alfv\'enic.  KHI can be still survived in the twisted tubes in nonmagnetic environment as the harmonics with sufficiently large $m$ are unstable for any sub-Alfv\'enic flow along the tubes (Zaqarashvili et al.\ 2010).

Magnetic structures observed in the solar wind (Bruno et al.\ \cite{Bruno2001}, Borovsky \cite{Borovsky2008}), which are believed to be magnetic flux tubes transported from the solar surface, may retain ``fossil'' properties typical to near-Sun conditions. Magnetic tubes near the solar photosphere, chromosphere and corona can be twisted due to various reasons: during raising phase along the convection zone (Moreno-Insertis and Emonet \cite{Moreno-Insertis1996}, Archontis et al.\ \cite{Archontis2004}, Murray and Hood \cite{Murray2008}, Hood et al.\ \cite{Hood2009}), by sunspot rotations (Khutsishvili et al.\ \cite{Khutsishvili1998}, Brown et al.\ \cite{Brown2003}, Yan and Qu \cite{Yan2007}, Zhang et al.\ \cite{Zhang2007}), and/or by magnetic tornadoes in the chromosphere and the corona (Wedemeyer-B\"ohm et al.\ \cite{Wedemeyer2012}, Su et al.\ \cite{Su2012}, Li et al.\ \cite{Li2012}). Solar prominences are also supposed to be formed in a twisted magnetic field (Priest et al.\ \cite{Priest1989}).  Therefore, magnetic flux tubes in the solar wind should be also twisted, which can be detected in situ observations as a variation of the total pressure (Zaqarashvili et al.\ \cite{Zaqarashvili2013}).

However, only the twist inside the magnetic tube is not sufficient for KHI of sub-Alfv\'enic motions as external flow-aligned magnetic field may stabilizes KHI when the vortices start to stretch the magnetic field lines outside the tube.  The solar wind magnetic field is generally directed along the Parker spiral, but individual magnetic tubes may move with an angle to the spiral. So that the external magnetic field is not necessarily directed along the motion of tube.  Therefore, the configuration of external magnetic field with regards to the tube motion can be extremely important for the KHI.

Here we studied KHI instability of twisted magnetic flux tube moving along its axis in the case of two different configurations of the external magnetic field.  First we assumed that the external magnetic field is directed along the tube axis, so that it is flow-aligned.  Then we assumed that the external magnetic field has a small twist, so that the external field has a small angle with the direction of the tube motion.  Both, tube and external magnetic fields are only slightly twisted, therefore they are stable against the kink instability.  We solved the incompressible MHD equations in cylindrical coordinates inside and outside the tube and obtained the transcendental dispersion equations through boundary conditions at the tube surface.  Then we solved the dispersion equations analytically in thin flux tube approximation and obtained the instability criteria for both untwisted and twisted external fields (Eqs.~(\ref{KH1}) and (\ref{KH2}), respectively).  We also solved the dispersion equations numerically and found the conditions for KHI.  Both, analytical and numerical solutions show that the KHI is suppressed for sub-Alfv\'enic motions when the twisted tubes moves along the Parker spiral, i.e., the external magnetic field is parallel to tube axis and the direction of motion.  So our results agree with already known scenario that the flow-aligned magnetic field stabilizes KHI.  However, if the external magnetic field has even very small twist, then the situation is completely changed.  We found that the harmonics satisfying the relation $\vec k \cdot \vec B \approx 0$ are unstable for any value of the flow.  The harmonics with higher $m$ are unstable for sub-Alfv\'enic motions.  The harmonics with wave vectors perpendicular to the magnetic field are in fact pure vortices in the incompressible limit.  Therefore, their instability to KHI has a real physical ground as they do not stretch significantly magnetic field lines.

Thus the analytical and numerical analyzes showed that the twisted tubes are always unstable to the KHI when they move in external twisted magnetic field.  This means that the magnetic tubes in the solar wind may excite Kelvin-Helmholtz vortices near boundaries.  These vortices may be responsible for initial energy in nonlinear cascade leading to MHD turbulence in the solar wind.

\section{Conclusions}

Twisted magnetic flux tubes can be unstable to KHI when they move with regards to the solar wind stream.  External axial magnetic field stabilizes KHI, therefore, the tubes moving along Parker spiral are unstable only for super-Alfv\'enic motions.  The instability criterion is
\[
|m|\mach^2> \left (1 + {\rho_\mathrm{i}\over \rho_\mathrm{e}}\right )\left (|m|{{B^2_{\mathrm{e} z}}\over {B^2_{\mathrm{i}z}}} + 1\right ).
\]
However, even a slight twist in the external magnetic field leads to KHI for any sub-Alfv\'enic motion.  Instability criterion in this case is
\[
|m|\mach^2> 1 + {{2+|m|}\over {|m|}}{\rho_\mathrm{i}\over \rho_\mathrm{e}},
\]
which shows that the modes with sufficiently large $m$ are always unstable for any value of the Alfv\'en Mach number.  The unstable harmonics satisfy the relation $\vec k \cdot \vec B \approx 0$, which corresponds to pure vortices in the incompressible MHD. Therefore, the twisted magnetic tubes moving with an angle to the Parker spiral may excite Kelvin-Helmholtz vortices, which may significantly contribute into the solar wind turbulence.

\begin{acknowledgements}
The work was supported by EU collaborative project STORM - 313038. The work of TZ was also supported by FP7-PEOPLE-2010-IRSES-269299 project- SOLSPANET, by Shota Rustaveli National Science Foundation grant DI/14/6-310/12 and by the Austrian "Fonds zur F\"{o}rderung der wissenschaftlichen Forschung" under projects P25640-N27 and P26181-N27.
The work of ZV was also supported by the Austrian Fonds zur F\"{o}rderung der wissenschaftlichen Forschung under project P24740-N27.  The work of IZh was supported by the Bulgarian Science Fund under project CSTC/INDIA 05/7.
\end{acknowledgements}

\end{document}